\documentclass[11pt,onecolumn]{IEEEtran}
\usepackage[latin1]{inputenc}
\usepackage{amsmath}
\usepackage{amsbsy}
\usepackage{amssymb}
\usepackage{latexsym}
\usepackage{times}
\usepackage{epsfig}
\title{Reduced-Rank Adaptive Filtering Based on Joint Iterative Optimization of Adaptive Filters}

\author{Rodrigo C. de Lamare, Member IEEE and Raimundo Sampaio-Neto  \\
\thanks{EDICS: SAS-ADAP. Dr. R. C. de Lamare is with the
Communications Research Group, Department of Electronics, University
of York, York Y010 5DD, United Kingdom and Prof. R. Sampaio-Neto is
with CETUC/PUC-RIO, 22453-900, Rio de Janeiro, Brazil. E-mails:
rcdl500@ohm.york.ac.uk, raimundo@cetuc.puc-rio.br} }

\begin{document}

\maketitle

\begin{abstract}
This letter proposes a novel adaptive reduced-rank filtering
scheme based on joint iterative optimization of adaptive filters.
The novel scheme consists of a joint iterative optimization of a
bank of full-rank adaptive filters that forms the projection
matrix and an adaptive reduced-rank filter that operates at the
output of the bank of filters. We describe minimum mean-squared
error (MMSE) expressions for the design of the projection matrix
and the reduced-rank filter and low-complexity normalized
least-mean squares (NLMS) adaptive algorithms for its efficient
implementation. Simulations for an interference suppression
application show that the proposed scheme outperforms in
convergence and tracking the state-of-the-art reduced-rank schemes
at significantly lower complexity.

\end{abstract}
\begin{keywords}
{Adaptive filters, iterative methods.}
\end{keywords}

\section{Introduction}

In adaptive filtering \cite{diniz}, one can find a huge number of
algorithms with different trade-offs between performance and
complexity. They range from the simple and low-complexity least-mean
squares (LMS) algorithms to the fast converging though complex
recursive least squares (RLS) techniques. Several attempts to
provide cost-effective adaptive filters with fast convergence
performance have been made with variable step-size algorithms,
data-reusing, sub-band and frequency-domain schemes and RLS
algorithms with linear complexity. A challenging problem which
remains unsolved by conventional techniques is that when the number
of elements in the filter is large, the algorithm requires a large
number of samples to reach its steady-state behavior. In these
situations, even RLS algorithms require an amount of data
proportional to $2M$ \cite{diniz} in stationary environments to
reach steady state, where $M$ is the filter length, and this may
lead to unacceptable convergence performance. In dynamic scenarios,
large filters usually fail or provide poor performance in tracking
signals embedded in interference.

Reduced-rank filtering \cite{scharf}-\cite{delamaresp} is a
powerful and effective technique in low sample support situations
and in problems with large filters. The advantages of reduced-rank
adaptive filters are their faster convergence speed and better
tracking performance than full-rank techniques when dealing with
large number of weights. Several reduced-rank methods and systems
have been proposed in the last several years, namely,
eigen-decomposition techniques \cite{bar-ness}-\cite{song&roy},
the multistage Wiener filter (MWF) \cite{gold&reed,goldstein} and
the auxiliary vector filtering (AVF) algorithm \cite{avf}. The
main problem with the above techniques is their high complexity
and the existence of numerical problems for implementation.

In this work we propose an adaptive reduced-rank filtering scheme
based on combinations of adaptive filters. Unlike related work on
combinations of full-rank filters \cite{garcia}, the novel scheme
consists of a joint iterative optimization of a bank of full-rank
adaptive filters which constitutes the projection matrix and an
adaptive reduced-rank filter that operates at the output of the
bank of full-rank filters. Differently from \cite{hua}, the
proposed scheme estimates a scalar, allows filter updates for each
successive observation, is adaptive and has low complexity. The
essence of the proposed approach is to change the role of adaptive
filters. The bank of adaptive filters is responsible for
performing dimensionality reduction, whereas the reduced-rank
filter effectively estimates the desired signal. Despite the large
dimensionality of the projection matrix and its associated slow
learning behavior, the proposed and existing \cite{goldstein,avf}
reduced-rank techniques enjoy in practice a very fast convergence.
The reason is that even an inaccurate or rough estimation of the
projection matrix is able to provide an appropriate dimensionality
reduction for the reduced-rank filter, whose behavior will govern
most of the performance of the overall scheme. We describe MMSE
expressions for the design of the projection matrix and the
reduced-rank filter along with simple NLMS adaptive algorithms for
its computationally efficient implementation. The performance of
the proposed scheme is assessed via simulations for CDMA
interference suppression.


\section{Reduced-Rank MMSE Parameter Estimation and Problem Statement}

The MMSE filter is the vector ${\bf w}= [w_1^{}~ w_2^{} ~ \ldots ~
w_M^{}]^T$, which is designed to minimize the MSE cost function
\begin{equation}
J = E\big[ | d(i) - {\bf w}^{H}{\bf r}(i)|^2 \big]
\end{equation}
where $d(i)$ is the desired signal, ${\bf r}(i)=[r_{0}^{(i)}~ \ldots
~r_{M-1}^{(i)}]^{T}$ is the received data, $(\cdot)^{T}$ and
$(\cdot)^{H}$ denote transpose and Hermitian transpose,
respectively, and $E[\cdot]$ stands for expectation. The set of
parameters ${\bf w}$ can be estimated via standard stochastic
gradient or least-squares estimation techniques \cite{diniz}.
However, the laws that govern the convergence behavior of these
estimation techniques imply that the convergence speed of these
algorithms is proportional to $M$, the number of elements in the
estimator. Thus, large $M$ implies slow convergence. A reduced-rank
algorithm attempts to circumvent this limitation in terms of speed
of convergence by reducing the number of adaptive coefficients and
extracting the most important features of the processed data. This
dimensionality reduction is accomplished by projecting the received
vectors onto a lower dimensional subspace. Specifically, consider an
$M \times D$ projection matrix ${\bf S}_{D}$ which carries out a
dimensionality reduction on the received data as given by
\begin{equation}
\bar{\bf r}(i) = {\bf S}_D^H {\bf r}(i)
\end{equation}
where, in what follows, all $D$-dimensional quantities are denoted
with a "bar". The resulting projected received vector $\bar{\bf
r}(i)$ is the input to a tapped-delay line filter represented by the
$D \time 1$ vector $\bar{\bf w}=[ \bar{w}_1^{}
~\bar{w}_2^{}~\ldots\bar{w}_D^{}]^T$ for time interval $i$. The
filter output corresponding to the $i$th time instant is
\begin{equation}
x(i) = \bar{\bf w}^{H}\bar{\bf r}(i)
\end{equation}
If we consider the MMSE design in (1) with the reduced-rank
parameters we obtain
\begin{equation}
\bar{\bf w} = \bar{\bf R}^{-1}\bar{\bf p}
\end{equation}
where $\bar{\bf R} = E[ \bar{\bf r}(i)\bar{\bf r}^{H}(i)]={\bf
S}_D^H{\bf R}{\bf S}_D$ is the reduced-rank covariance matrix, ${\bf
R} = E[{\bf r}(i){\bf r}^{H}(i)]$ is the full-rank covariance
matrix, $\bar{\bf p}=E[d^*(i)\bar{\bf r}(i)]={\bf S}_D^H{\bf p}$ and
${\bf p}=E[d^*(i){\bf r}(i)]$. The associated MMSE for a rank $D$
estimator is expressed by
\begin{equation}
{\rm MMSE} = \sigma^2_d - \bar{\bf p}^H \bar{\bf R}^{-1}\bar{\bf p}
= \sigma^2_d - {\bf p}^H{\bf S}_D ({\bf S}_D^H{\bf R}{\bf S}_D)^{-1}
{\bf S}_D^H{\bf p}
\end{equation}
where $\sigma^2_d$ is the variance of $d(i)$. In the Appendix, we
provide a necessary and sufficient condition for a projection
${\bf S}_D$ with dimensions $M \times D$ to not modify the MMSE
and discuss the existence of multiple solutions. Based upon the
problem statement above, the rationale for reduced-rank schemes
can be simply put as follows. How to efficiently (or optimally)
design a transformation matrix ${\bf S}_D$ with dimensions $M
\times D$ that projects the observed data vector ${\bf r}(i)$ with
dimensions $M \times 1$ onto a reduced-rank data vector $\bar{\bf
r}(i)$ with dimensions $D \times 1$? In the next section we
present the proposed reduced-rank approach.

\section{Proposed Reduced-Rank Scheme}

Here we detail the principles of the proposed reduced-rank scheme
using a projection operator based on adaptive filters. The novel
scheme, depicted in Fig. 1, employs a projection matrix ${\bf
S}_{D}(i)$ with dimensions $M \times D$ to process a data vector
with dimensions $M \times 1$, that is responsible for the
dimensionality reduction. The reduced-rank filter $\bar{\bf w}(i)$
with dimensions $D \times 1$ processes the reduced-rank data
vector $\bar{\bf r}(i)$ in order to yield a scalar estimate
$x(i)$. The projection matrix ${\bf S}_{D}(i)$ and the
reduced-rank filter $\bar{\bf w}(i)$ are jointly optimized in the
proposed scheme according to the MMSE criterion. \vspace*{-1em}

\begin{figure}[htb]
       \centering  
       \hspace*{-2.75em}
        \vspace*{-2em}
      {\includegraphics[width=11cm, height=4.0cm]{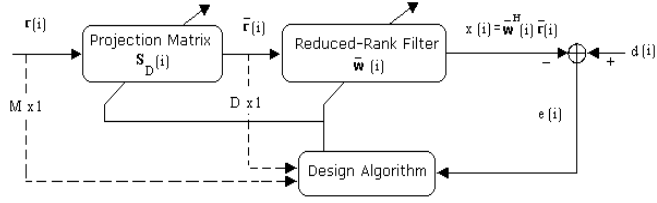}}
       \vspace*{-1em}
       \caption{\small Proposed Reduced-Rank Scheme.}
\end{figure}

Specifically, the projection matrix is structured as a bank of $D$
full-rank filters ${\bf s}_d(i)=[s_{1,d}(i) ~ s_{2,d}(i)~
\ldots~s_{M,d}(i) ]^T$ ($d = 1,~\ldots,~D$) with dimensions $M
\times 1$ as given by ${\bf S}_{D}(i) = [~{\bf s}_1(i) ~| ~{\bf
s}_2(i)~| ~\ldots~|{\bf s}_D(i)~]$. Let us now mathematically
express the output estimate $x(i)$ of the reduced-rank scheme as a
function of the received data ${\bf r}(i)$, the projection matrix
${\bf S}_D(i)$ and the reduced-rank filter $\bar{\bf w}(i)$:
\begin{equation}
\begin{split}
x(i) & =  \bar{\bf w}^H(i) {\bf S}_D^H(i) {\bf r}(i) = \bar{\bf
w}^H(i) \bar{\bf r}(i)
\end{split}
\end{equation}
Note that for $D=1$, the novel scheme becomes a conventional
full-rank filtering scheme with an addition weight parameter $w_D$
that provides a gain. For $D>1$, the signal processing tasks are
changed and the full-rank filters compute a subspace projection and
the reduced-rank filter estimates the desired signal.

The MMSE expressions for the filters ${\bf S}_D(i)$ and $\bar{\bf
w}(i)$ can be computed through the following cost function:
\begin{equation}
\begin{split}
J & = E\big[ | d(i) - \bar{\bf w}^{H}(i){\bf S}_D^H(i){\bf
r}(i)|^2 \big] \\
& = E\big[ | d(i) - \bar{\bf w}^{H}(i)\bar{\bf r}(i)|^2 \big]
\end{split}
\end{equation}
By fixing the projection ${\bf S}_D(i)$ and minimizing (7) with
respect to $\bar{\bf w}(i)$, the reduced-rank filter weight vector
becomes
\begin{equation}
\bar{\bf w}(i) = \bar{\bf R}^{-1}(i) \bar{\bf p}(i)
\end{equation}
where $\bar{\bf R}(i) = E[{\bf S}_D^H(i){\bf r}(i){\bf r}^H(i) {\bf
S}_D(i) ] =E[\bar{\bf r}(i) \bar{\bf r}^{H}(i)]$, $\bar{\bf p}(i) =
E[d^{*}(i){\bf S}_D^H(i){\bf r}(i)] = E[d^{*}(i)\bar{\bf r}(i)]$. We
proceed with the proposed joint optimization  by fixing $\bar{\bf
w}(i)$ and minimizing (7) with respect to ${\bf S}_D(i)$. We then
arrive at the following expression for the projection operator
\begin{equation} {\bf S}_D(i) = {\bf R}^{-1}(i) {\bf P}_D(i) {\bf
R}_{w}(i)
\end{equation}
where  ${\bf R}(i) = E[{\bf r}(i){\bf r}^{H}(i)]$, ${\bf P}_D(i) =
E[d^{*}(i){\bf r}(i){\bf w}^H(i)]$ and ${\bf R}_w(i) = E[{\bf
w}(i){\bf w}^{H}(i)]$. The associated MMSE is
\begin{equation}
{\rm MMSE} = \sigma^{2}_{d} - \bar{\bf p}^{H}(i) \bar{\bf R}^{-1}(i)
\bar{\bf p}(i)
\end{equation}
where $\sigma^{2}_{d}=E[|d(i)|^{2}]$. Note that the filter
expressions in (8) and (9) are not closed-form solutions for
$\bar{\bf w}(i)$ and ${\bf S}_D(i)$ since (8) is a function of ${\bf
S}_D(i)$ and (9) depends on $\bar{\bf w}(i)$ and thus it is
necessary to iterate (8) and (9) with an initial guess to obtain a
solution. The MWF \cite{gold&reed} employs the operator ${\bf S}_D =
\big[{\bf p}~{\bf R}{\bf p}~\ldots~ {\bf R}^{D-1}{\bf p}\big]$ that
projects the data onto the Krylov subspace. Unlike the MWF approach,
the new scheme provides an iterative exchange of information between
the reduced-rank filter and the projection matrix and leads to a
much simpler adaptive implementation than the MWF. The projection
matrix reduces the dimension of the input data, whereas the
reduced-rank filter attempts to estimate the desired signal. The key
strategy lies in the joint optimization of the filters. The rank $D$
must be set by the designer to ensure appropriate performance and
the reader is referred to \cite{qian} for rank selection methods. In
the next section, we seek iterative solutions via adaptive
algorithms.

\section{Adaptive Algorithms}
\label{sec:typestyle}

Here we describe an adaptive NLMS implementation, convergence
conditions and detail the computational complexity in arithmetic
operations of the proposed reduced-rank scheme.

\subsection{Adaptive Algorithms}

Let us consider the following Lagrangian cost function
\begin{equation}
\begin{split}
{\mathcal{L}}  & = ||{\bf w}(i+1) - {\bf w}(i)||^2 + ||{\bf
S}_D(i+1) - {\bf S}_D(i)||^2
\\ & \quad + \Re [\lambda_1^* (d(i) - {\bf w}^H(i+1){\bf S}_D^H(i){\bf
r}(i) ]  \\
& \quad + \Re [\lambda_2^* (d(i) - {\bf w}^H(i){\bf S}_D^H(i+1){\bf
r}(i) ],
\end{split}
\end{equation}
where $\lambda_1$, $\lambda_2$ are scalar Lagrange multipliers,
$||\cdot ||^2$ denotes the Frobenius norm and the operator
$\Re[\cdot]$ retains the real part of the argument. By computing
the gradient terms of (11) with respect to $\bar{\bf w}(i+1)$,
${\bf S}_D(i+1)$, $\lambda_1$ and $\lambda_2$, setting them to $0$
and solving the resulting equations, we obtain:
\begin{equation}
\nabla_{\bar{\bf w}(i+1)} {\mathcal{L}}  = 2(\bar{\bf w}(i+1) -
\bar{\bf w}(i)) + {\bf S}_D^H(i){\bf r}(i) \lambda_1={\bf 0}
\end{equation}
\begin{equation}
\nabla_{{\bf S}_D(i+1)} {\mathcal{L}} = 2({\bf S}_D(i+1) - {\bf
S}_D(i)) + {\bf r}(i)\bar{\bf w}^H(i)\lambda_2={\bf 0}
\end{equation}
\begin{equation}
\nabla_{\lambda_1} {\mathcal{L}} = d(i) - \bar{\bf w}^H(i+1){\bf
S}_D^H(i){\bf r}(i) ={ 0}
\end{equation}
\begin{equation}
\nabla_{\lambda_2} {\mathcal{L}} = d(i) - \bar{\bf w}^H(i){\bf
S}_D^H(i+1){\bf r}(i) ={ 0}
\end{equation}
By solving the above equations and introducing the convergence
factors $\mu_0$ and $\eta_0$, the proposed jointly optimized and
iterative NLMS algorithms for parameter estimation become
\begin{equation}
\bar{\bf w}(i+1) = \bar{\bf w}(i) + \mu(i) e^*(i)\bar{\bf r}(i),
\end{equation}
\begin{equation}
{\bf S}_D(i+1) = {\bf S}_D(i) + \eta(i) e^*(i){\bf r}(i)\bar{\bf
w}^H(i),
\end{equation}
where $e(i)= d(i) - \bar{\bf w}^H(i){\bf S}_D^H(i){\bf r}(i)$,
$\mu(i)= \mu_0/({\bf r}^H(i){\bf r}(i)) $ and $\eta(i)=
\eta_0/(\bar{\bf w}^H(i)\bar{\bf w}(i){\bf r}^H(i){\bf r}(i))$ are
the time-varying step sizes. The algorithms described in (16)-(17)
have a complexity $O(DM)$. The proposed scheme trades-off a
full-rank filter against $D$ full-rank adaptive filters as the
projection matrix ${\bf S}_D(i)$ and one reduced-rank adaptive
filter $\bar{\bf w}(i)$ operating simultaneously and exchanging
information. The iteration and convergence occurs over several
observations and here we consider only one iteration per symbol
$(i)$.

\subsection{Convergence Conditions}

Define the error matrices at time index $i$ as ${\bf e}_{\bar{\bf
w}}(i) = \bar{\bf w}(i) - \bar{\bf w}_{\rm opt}$ and ${\bf e}_{{\bf
S}_D}(i) = {\bf S}_D(i) - {\bf S}_{D,{\rm opt}}$, where $\bar{\bf
w}_{\rm opt}$ and ${\bf S}_{D,{\rm opt}}$ are the optimal parameter
estimators. Because of the joint optimization procedure, both
filters have to be considered jointly. By substituting the
expressions of ${\bf e}_{\bar{\bf w}}(i)$ and ${\bf e}_{{\bf
S}_D}(i)$ in (16) and (17), taking expectations and simplifying the
terms, we obtain
\begin{equation}
\begin{split}
\left[\begin{array}{c}
  E[{\bf e}_{\bar{\bf w}}(i+1)] \\
  E[{\bf e}_{{\bf S}_D}(i+1)]
\end{array}\right] & = {\bf A}
\left[\begin{array}{c}
  E[{\bf e}_{\bar{\bf w}}(i)] \\
  E[{\bf e}_{{\bf S}_D}(i)]
\end{array}\right] + {\bf B} \\
 \end{split}
\end{equation}
where ${\bf A} =  \left[\begin{array}{c c}
  {\small ({\bf I} - E[\mu(i)] \bar{\bf R})} & {\small {\bf 0}} \\
  {\small E[\nu(i)] \sigma_w^2 {\bf R} {\bf S}_{D,{\rm opt}}} & {\small ({\bf I} -
  E[\nu(i)]\sigma_w^2 {\bf R})}
\end{array}\right]$,
${\bf B} = \left[\begin{array}{c}
{\small E[\mu(i)] ({\bf R}{\bf S}_D(i)\bar{\bf w}_{\rm opt} - \bar{\bf p})} \\
{\small E[\nu(i)] \sigma_w^2( {\bf R}{\bf S}_{D,{\rm opt}}
\bar{\bf w}_{\rm opt} - {\bf p}) } \end{array}\right]$ and
$\sigma_w^2 = E[ ||\bar{\bf w}(i)||^2]$. The above equation
implies that the stability of the algorithms depends on the
spectral radius of ${\bf A}$. For convergence, the step sizes
should be chosen such that the eigenvalues of ${\bf A}^H{\bf A}$
are less than one.

\subsection{Computational Complexity}

Here, we detail the computational complexity in terms of additions
and multiplications of the proposed schemes with NLMS and other
existing algorithms, namely the Full-rank with NLMS and RLS, the
MWF \cite{goldstein} with NLMS and RLS and the AVF \cite{avf}, as
shown in Table 1. The MWF \cite{goldstein} has a complexity $O(D
\bar{M}^{2})$, where the variable dimension of the vectors
$\bar{M} = M - d$ varies according to the rank $d = 1, \ldots, D$.
The proposed scheme is much simpler than the Full-rank with RLS,
the MWF and the AVF and slightly more complex than the Full-rank
with NLMS (for $D << M$, as will be explained later).

\vspace*{-1.0em}{
\begin{table}[h]
\centering%
\caption{\small Computational complexity of algorithms.}
\vspace*{-1.0em}{
\begin{tabular}{lcc}
\hline \rule{0cm}{2.0ex}&  \multicolumn{2}{c}{\small Number of
operations per symbol } \\ \cline{2-3}
{\small Algorithm} & {\small Additions} & {\small Multiplications} \\
\hline
\emph{\small \bf Full-rank-NLMS} & {\footnotesize $3M-1$} & {\footnotesize $3M+2$}  \\
\emph{\small \bf Full-rank-RLS} & {\footnotesize $3(M-1)^{2} + M^{2} + 2M$} & {\footnotesize $6M^{2}+2M + 2$}   \\
\emph{\small \bf Proposed-NLMS}   & {\footnotesize $2DM+M+4D-2$} & {\footnotesize $ 3DM+M+3D+6$}  \\
\emph{\small \bf MWF-NLMS} & {\footnotesize $D(2\bar{M}^{2} -3\bar{M} + 1)$} & {\footnotesize $D(2\bar{M}^{2} +5\bar{M} + 7)$}  \\
\emph{\small \bf MWF-RLS}  & {\footnotesize $D(4(\bar{M}-1)^{2} + 2\bar{M})$} & {\footnotesize $D(4\bar{M}^{2} +2\bar{M} + 3)$} \\
\emph{\small \bf AVF} & {\footnotesize $D(M^2+3(M-1)^2)-1$} & {\footnotesize $D(4M^2+4M + 1)$} \\
\emph{\small } & {\footnotesize $+D(5(M-1)+1)+2M$} & {\footnotesize $+ 4M + 2$} \\
\hline
\end{tabular}
}
\end{table}}
\section{Simulations}

In this section we assess the proposed reduced-rank scheme and
algorithms in a CDMA interference suppression application. We
consider the uplink of a symbol synchronous BPSK DS-CDMA system
with $K$ users, $N$ chips per symbol and $L$ propagation paths.
Assuming that the channel is constant during each symbol interval
and the randomly generated spreading codes are repeated from
symbol to symbol, the received signal after filtering by a
chip-pulse matched filter and sampled at chip rate yields the
$M$-dimensional received vector
\begin{equation}
{\bf r}(i)  = \sum_{k=1}^{K} {\bf H}_{k}(i) A_{k} {\bf C}_{k}{\bf
b}_{k}(i)  + {\bf n}(i),
\end{equation}
where $M=N+L-1$, ${\bf n}(i) = [n_{1}(i) ~\ldots~n_{M}(i)]^{T}$ is
the complex Gaussian noise vector with $E[{\bf n}(i){\bf n}^{H}(i)]
= \sigma^{2}{\bf I}$, the symbol vector is ${\bf b}_{k}(i) =
[b_{k}(i+L_{s}-1)~\ldots ~ b_{k}(i)~\ldots ~ b_{k}(i-L_{s}+1)]^{T}$,
the amplitude of user $k$ is $A_{k}$, $L_{s}$ is the intersymbol
interference span, the $((2L_{s}-1) \cdot N)\times (2L_{s}-1)$ block
diagonal matrix ${\bf C}_{k}$ is formed with $N$-chips shifted
versions of the signature ${\bf s}_{k} = [a_{k}(1) \ldots
a_{k}(N)]^{T}$ of user $k$ and the $ M~\times (2\cdot L_{s}-1) \cdot
N$ convolution matrix ${\bf H}_k(i)$ is constructed with shifted
versions of the $L\times 1$ channel vector ${\bf h}_k(i) =
[{h}_{k,0}(i) ~\ldots ~ {h}_{k,L_{p}-1}(i)]^T$ on each column and
zeros elsewhere. For all simulations, we use $\bar{\bf w}(0)={\bf
0}_{D,1}$, ${\bf S}_D(0)=[{\bf I}_D ~  {\bf 0}_{D,M-D}]^T$, assume
$L=9$ as an upper bound, use $3$-path channels with relative powers
given by $0$, $-3$ and $-6$ dB, where in each run the spacing
between paths is obtained from a discrete uniform random variable
between $1$ and $2$ chips and average the experiments over $200$
runs. The system has a power distribution amongst the users for each
run that follows a log-normal distribution with associated standard
deviation equal to $1.5$ dB.

\begin{figure}[!htb]
\begin{center}
\def\epsfsize#1#2{0.625\columnwidth}
\epsfbox{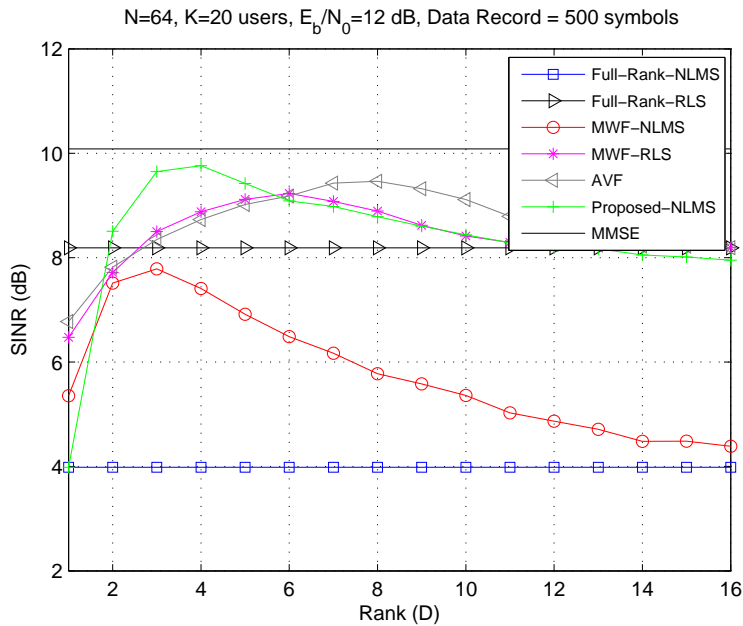} \vspace*{-1em}\caption{\small SINR
performance versus rank (D).}
\end{center}
\end{figure}

We compare the proposed scheme with the Full-rank \cite{diniz},
the MWF \cite{goldstein} and the AVF \cite{avf} techniques for the
design of linear receivers, where the reduced-rank filter
$\bar{\bf w}(i)$ with $D$ coefficients provides an estimate of the
desired symbol for the desired user (user $1$ in all experiments)
using the signal-to-interference-plus-noise ratio (SINR)
\cite{goldstein}. We consider the SINR performance versus the rank
$D$ with optimized parameters ($\mu_0$, $\nu_0$ and forgetting
factors $\lambda$) for all schemes. The results in Fig. 2 indicate
that the best rank for the proposed scheme is $D=4$ (which will be
used in the remaining experiments) and it is very close to the
optimal full-rank MMSE. Studies with systems with different
processing gains show that $D$ is invariant to the system size,
which brings considerable computational savings. In practice, the
rank $D$ can be adapted in order to obtain fast convergence and
ensure good steady state performance and tracking after
convergence.

We show an experiment in Fig. 3 where the adaptive filters are set
to converge to the same SINR. The NLMS version of the MWF is known
to have problems in these situations since it does not
tridiagonalize its covariance matrix \cite{goldstein} and thus is
unable to approach the MMSE. The curves show an excellent
performance for the proposed scheme and algorithms, which converge
much faster than the full-rank filter, are comparable to the more
complex MWF-RLS and AVF schemes, at much lower complexity.

\begin{figure}[!htb]
\begin{center}
\def\epsfsize#1#2{0.625\columnwidth}
\epsfbox{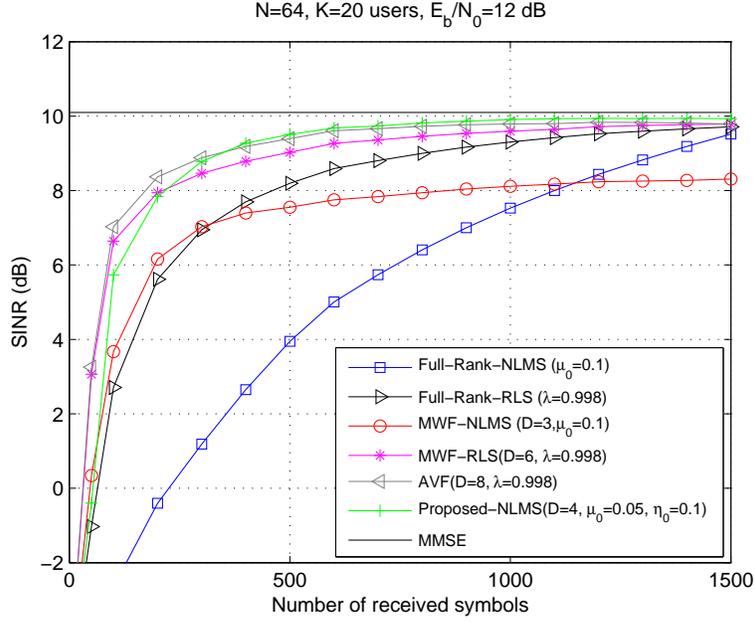} \vspace*{-1em}\caption{\small SINR
performance versus number of received symbols. }
\end{center}
\end{figure}

The BER convergence performance in a mobile communications situation
is shown in Fig. 4. The channel coefficients are obtained with
Clarke´s model \cite{rappa} and the adaptive filters of all methods
are trained with $250$ symbols and then switch to decision-directed
mode. The results show that the proposed scheme has a much better
performance than the existing approaches and is able to adequately
track the desired signal. 

\begin{figure}[!htb]
\begin{center}
\def\epsfsize#1#2{0.625\columnwidth}
\epsfbox{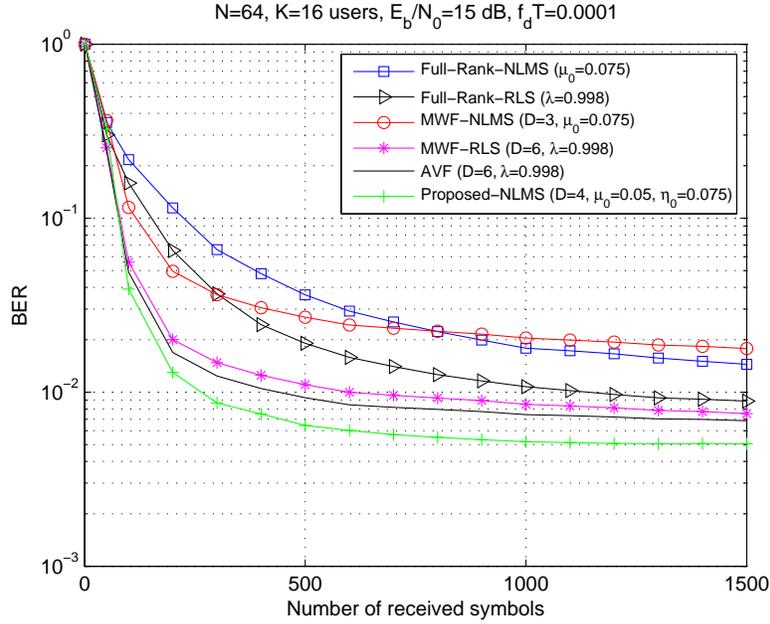} \vspace*{-1em}\caption{\small BER
performance versus number of received symbols.}
\end{center}
\end{figure}

\section{Conclusions}

We proposed a novel reduced-rank scheme based on joint iterative
optimization of adaptive filters with a low complexity
implementation using NLMS algorithms. In the proposed scheme, the
full-rank adaptive filters are responsible for estimating the
subspace projection rather than the desired signal, which is
estimated by a small reduced-rank filter. The results for CDMA
interference suppression show a performance significantly better
than existing schemes and close to the optimal full-rank MMSE.

\begin{appendix}

Given a $M \times D$ projection matrix ${\bf S}_D$, where $D \leq
M$, the ${\rm MMSE}$ is achieved if and only if ${\bf w}$ which
minimizes (1) belongs to the ${\rm Range}\{ {\bf S}_D \}$, i.e.
${\bf w}$ lies in the subspace generated by ${\bf S}_D$. In this
case, we have ${\rm MMSE}(\bar{\bf w}) = {\rm MMSE}({\bf
w})=\sigma_d^2 - {\bf p}^H {\bf R}^{-1}{\bf p}$. For a general ${\bf
S}_D$, we have ${\rm MMSE}(\bar{\bf w}) \geq \sigma_d^2 - {\bf p}^H
{\bf R}^{-1}{\bf p}$. From the above analysis, we can conclude that
there exists multiple solutions to the proposed optimization
problem. However, our studies indicate that there are no local
minima and the performance is insensitive to initialization,
provided we select the initial values $\bar{\bf w}(0)$ and ${\bf
S}_D(0)$ which do not instabilize the algorithm and annihilate the
signal, respectively.

\end{appendix}

\end{document}